\documentclass[aps,superscriptaddress]{revtex4}
\usepackage{graphicx}

\begin{document}

\title{Optical hollow-core waves in nonlinear Epsilon-Near-Zero metamaterials}

\author{C. Rizza}
\email{carlo.rizza@aquila.infn.it} \affiliation{Dipartimento di Ingegneria Elettrica e dell'Informazione, Universit$\grave{a}$ dell'Aquila, 67100
L'Aquila, Italy} \affiliation{Consiglio Nazionale delle Ricerche, CNR-SPIN 67100 L'Aquila, Italy and Dipartimento di Fisica, Universit$\grave{a}$
dell'Aquila, 67100 L'Aquila, Italy}

\author{A. Ciattoni}
\affiliation{Consiglio Nazionale delle Ricerche, CNR-SPIN 67100 L'Aquila, Italy and Dipartimento di Fisica, Universit$\grave{a}$ dell'Aquila, 67100
L'Aquila, Italy}

\author{E. Palange}
\affiliation{Dipartimento di Ingegneria Elettrica e dell'Informazione, Universit$\grave{a}$ dell'Aquila, 67100 L'Aquila, Italy}
\date{\today}

\begin{abstract}
We investigate non-diffracting hollow-core nonlinear optical waves propagating in a layered nanoscaled metal-dielectric structure characterized by a very
small average linear dielectric permittivity (Nonlinear Epsilon-Near-Zero metamaterial). We analytically show that hollow-core waves have a power flow
exactly vanishing at a central region and exhibiting a sharp sloped profile at the edges of the regions surrounding the core. Physically, the absence of
power flow at the core region is due to the vanishing of the transverse component of the electric displacement field, condition that can be satisfied by
the full compensation between the nonlinear and linear dielectric contribution.
\end{abstract}

\maketitle

\section{Introduction}

Electromagnetic metamaterials have recently attracted significant research attention due to their unique properties generally not available in
standard media. The studies of linear metamaterials have opened new possibilities to achieve remarkable effects like as, for example, superlensing
\cite{Pendry1}, optical cloaking \cite{Pendry2},nano-photonic circuits \cite{Engheta1,Engheta2}. In the same way, the research on the nonlinear
metamaterials can overcome the limits of the naturally-occurring materials and can expand the range of possible applications
\cite{Pendry3,Zharov,Liu,Powell,Ciatt3}. As a leading example, the artificial fabrication of nonlinear materials suggested the way to enhanced the
nonlinearity at low intensity. Pendry et al., in their pioneering work \cite{Pendry3}, proposed a left-handed metamaterials characterized by a huge
nonlinear response arising from the concentration (due to microscopic inhomogeneity) of the electromagnetic field within the nonlinear constituents.
On the other hand, a simple strategy for making available the nonlinearity is to reduce the linear polarization. Following this strategy, Ciattoni et
al. \cite{Ciatt1} suggested a simple composite structure to achieve very small value of dielectric permittivity where the nonlinear contribution to
the dielectric response is not a mere perturbation of the linear one (extreme nonlinear regime). Authors considered a nanoscaled layered structure
consisting of alternating metal and dielectric slabs and, in the effective medium scheme, characterized by effective constitutive relations formally
coinciding with those of a standard Kerr medium and with response parameters assuming values not available in standard media (nonlinear
metamaterials). In these kind of metamaterials, transverse magnetic nonlinear guided waves show a very rich and exotic phenomenology such as the
transverse power flow reversing (i.e. the power flow of the considered waves changes its sign along the wave transverse profile, see also
Ref.\cite{Ciatt2}).
\begin{figure}[htbp]
\begin{center}
\includegraphics[width=0.5\linewidth]{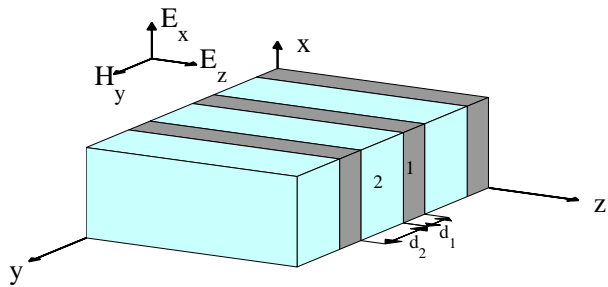}
\caption{Geometry of nanoscaled periodic structure and TM electromagnetic field configuration where the unit cell is composed by two layers ($N=2$).}
\label{fig1}
\end{center}
\end{figure}

In this paper, we consider specific transverse magnetic nonlinear guided waves propagating in a nonlinear Epsilon-Near-Zero metamaterial and
characterized by a power flow exactly vanishing in a core region and exhibiting a sharp sloped profile at the edges of the side regions surrounding
the core. The very marked steepening of the power flow and of the underlying electric field transverse component $E_x$ is a consequence of the
vanishing of the transverse component of the displacement field along the transverse wave profile, i.e. $D_x(x)=0$. Since $D_x$ is nonlinearly
related to $E_x$ and $E_z$, this requirement is differently fulfilled within the core (where $E_x=0$) and the lateral regions (where the effective
nonlinear dielectric permittivity vanishes) thus yielding a discontinuity of the transverse field derivatives at the core edges. Such an extreme
electromagnetic configuration has, as a main underlying physical mechanism, the exact compensation between linear and nonlinear contributions to the
dielectric response in the regions surrounding the core and this is possible since the considered medium can support the extreme nonlinear regime.

\section{Nonlinear Epsilon-Near-Zero Metamaterials}

Consider a monochromatic Transverse Magnetic (TM) field (of time dependence $\exp(-i\omega t)$) ${\bf E} = E_x(x,z) \hat{\bf e}_x+ E_z(x,z) \hat{\bf
e}_z$, ${\bf H} =  H_y(x,z) \hat{\bf e}_y$, propagating along the $z$-axis in a periodic layered medium. The unit cell (of spatial period $d$) is
composed by $N$ layers filled by different media of thickness $d_j$ (as reported in Fig.1 for the case $N=2$). We suppose that the $j$-th layer is
characterized by the constitutive relation
\begin{equation} \label{H-D_0}
{\bf D_j}= \epsilon_0 \epsilon_j {\bf E}_j+\epsilon_0 \chi_j \left[ |{\bf E}_j|^2 {\bf E}_j +\gamma_j ({\bf E}_j \cdot {\bf E}_j) {\bf E}_j^* \right],
\end{equation}
where ${\bf E}_j$,${\bf D}_j$ are the local electric and electric displacement field, respectively; whereas $\epsilon_j$ is the dielectric permittivity
of the $j$th layer and $\chi_j$ and $\gamma_j$ are the standard nonlinear Kerr parameters. Under the assumption that the light wavelength $\lambda$ is
much greater than the spatial period $d$, the homogenized medium behaves like a Kerr medium whose effective constitutive relation is
\begin{equation} \label{H-D}
{\bf D}= \epsilon_0 \epsilon {\bf E}+\epsilon_0 \chi \left[ |{\bf E}|^2 {\bf E} +\gamma ({\bf E} \cdot {\bf E}) {\bf E}^* \right],
\end{equation}
where ${\bf E}=\langle{\bf E}_j\rangle$,${\bf D}=\langle{\bf D}_j\rangle$, $\epsilon=\langle\epsilon_j\rangle$, $\chi=\langle\chi_j\rangle$ and
$\gamma=\langle\chi_j \gamma_j \rangle/\langle\chi_j \rangle$ (each field and parameter is obtained by averaging along the $y$-axis the layer local
quantities over the period $d$) \cite{Ciatt1}. All effective parameters appearing in the relation (\ref{H-D}) can be tailored by choosing the
underlying constituents; for example the averaging of the dielectric permittivity can be used to obtain an efficient loss management (using gain and
loss media as underlying layers) \cite{Pendry4} and/or an epsilon-near-zero metamaterials (exploiting negative and positive dielectric layers)
\cite{Engheta2}. It is worth noting that, the nonlinear parameter $\gamma$ can span the whole range of the real values whereas, in the standard Kerr
material, it can assume only three values \cite{Boyd} depending on the physical mechanism providing the Kerr nonlinear response.

\section{Hollow-core waves}

We focus here on a situation where $0<\epsilon \ll 1$, $\chi<0$ and $\gamma<-1$. Eliminating the magnetic field from Maxwell's equations $\nabla
\times {\bf E}=i \omega \mu_0 {\bf H}$, $\nabla \times {\bf H}=-i \omega {\bf D}$ and considering TM non-diffracting fields of the form
$E_x(\xi,\zeta) = \sqrt{\epsilon/|\chi|} u_x(\xi)\exp{(i \beta \zeta)}$, $E_z(\xi,\zeta) = i \sqrt{\epsilon/|\chi|} u_z(\xi)\exp{(i \beta\zeta)}$, we
obtain the system
\begin{eqnarray} \label{syst2}
&&-\beta \frac{d u_z}{d \xi}+ \left[\beta^2 -  \epsilon_x^{(NL)} \right] u_x =0, \nonumber\\
&&\beta \frac{d u_x}{d \xi}- \frac{d^2 u_z}{ d \xi^2} - \epsilon_z^{(NL)} u_z=0,
\end{eqnarray}
where $\beta$ is a real propagation constant, $u_x$ and $u_z$ are real field profiles, $\xi = \sqrt{\epsilon }k_0 x$, $\zeta = \sqrt{\epsilon } k_0
z$ are dimensionless variables and we defined
\begin{eqnarray} \label{epNL}
\epsilon_x^{(NL)} &=&  1-\left[ (1+\gamma) u_x^2+(1-\gamma)u_z^2  \right],\nonumber\\
\epsilon_z^{(NL)} &=& 1-\left[ (1-\gamma) u_x^2+(1+\gamma)u_z^2  \right].
\end{eqnarray}
as effective normalized nonlinear dielectric permittivities. The system of Eqs.(\ref{syst2}) is integrable since it admits the first integral $F =
\beta^2 u_x^2- \left[ \frac{du_z}{d\xi} \right]^2- (u_x^2+u_z^2) +\left[(1-\gamma)u_x^2u_z^2+\frac{1}{2}(1+\gamma)(u_x^4+u_z^4)\right]$, i.e.
$dF/d\xi=0$ is satisfied for any solution $u_x(\xi)$, $u_z(\xi)$ of Eqs.(\ref{syst2}). We focus on solutions where $u_x$ is spatially even
($u_x(\xi)=u_x(-\xi)$) and $u_z$ is odd ($u_z(\xi)=-u_z(-\xi)$) so that we assume the boundary conditions $u_x(\pm\infty)=u_{x\infty}$ and
$u_z(\pm\infty)=\pm u_{z\infty}$. Since $u_x$ and $u_z$ asymptotically approach to two constant values, we require their derivatives to
asymptotically vanish and, as a consequence, we obtain from Eqs.(\ref{syst2}) (evaluated for $\xi \rightarrow +\infty$)
\begin{eqnarray} \label{bet}
\beta^2 &=& 2 \gamma ( 1 - 2 u_{x\infty}^2 )/(1+\gamma) \nonumber \\
u_{z\infty}^2 &=& \left[1-(1-\gamma)u_{x\infty}^2 \right]/(1+\gamma).
\end{eqnarray}
Substituting this expression of $\beta$ and $du_z/d\xi$, derived from first of Eq.(\ref{syst2}), into the expression of $F$, we obtain $F$ as a
function of $u_x$ and $u_z$, so that the nonlinear guided waves correspond to curves of constant $F(u_x,u_z)$, in the plane $(u_x,u_z)$, which are
heteroclinic orbits joining the two saddle points $(u_{x\infty},\pm u_{z\infty})$. It can be shown that such nonlinear guided waves exist if
$1/(1-\gamma) < u_{x\infty}^2< 1/2$ (see Ref.\cite{Ciatt1} for a more detailed discussion). It is worth noting that Eqs.(\ref{syst2}) can be casted
into the form
\begin{eqnarray} \label{syst3}
&& \beta \frac{d u_z}{d \xi} =\left[\beta^2- \epsilon_x^{(NL)}\right] u_x, \nonumber\\
&& \beta \frac{d u_x}{d \xi} = \frac{\beta^2 \epsilon_z^{(NL)} u_z +\left( \epsilon_x^{(NL)} -\beta^2 \right)  \displaystyle \frac{\partial
\epsilon_x^{(NL)} }{\partial u_z} u_x^2 }{ \epsilon_x^{(NL)}+ \displaystyle \frac{\partial \epsilon_x^{(NL)}}{\partial u_x} u_x },
\end{eqnarray}
which is a system of the first order fully equivalent to Maxwell's equations with the two constraints
\begin{equation} \label{constraints}
\beta \neq 0, \quad \epsilon_x^{(NL)} + \displaystyle \frac{\partial \epsilon_x^{(NL)}}{\partial u_x} u_x  \neq 0.
\end{equation}
We have evaluated the profiles of $u_x(\xi)$ and $u_z(\xi)$ by solving Eqs.(\ref{syst3}) (with a standard numerical Runge-Kutta method) along with
the initial conditions $u_x(0)=u_{x0}$ (obtained from the first integral) and $u_z(0)=0$. In Fig.(2) we report the profiles of $u_x(\xi)$ and
$u_z(\xi)$ of various nonlinear guided waves (solid lines) and it is evident that, for $u_{x\infty}^2 \rightarrow 1/2$, the profile of $u_x(\xi)$
approaches a limiting shape (reported in dotted line) characterized by a core region where $u_x = 0$ and lateral regions at whose edges $u_x$
exhibits very large slopes.
\begin{figure}[htbp]
\begin{center}
\includegraphics[width=0.5\linewidth]{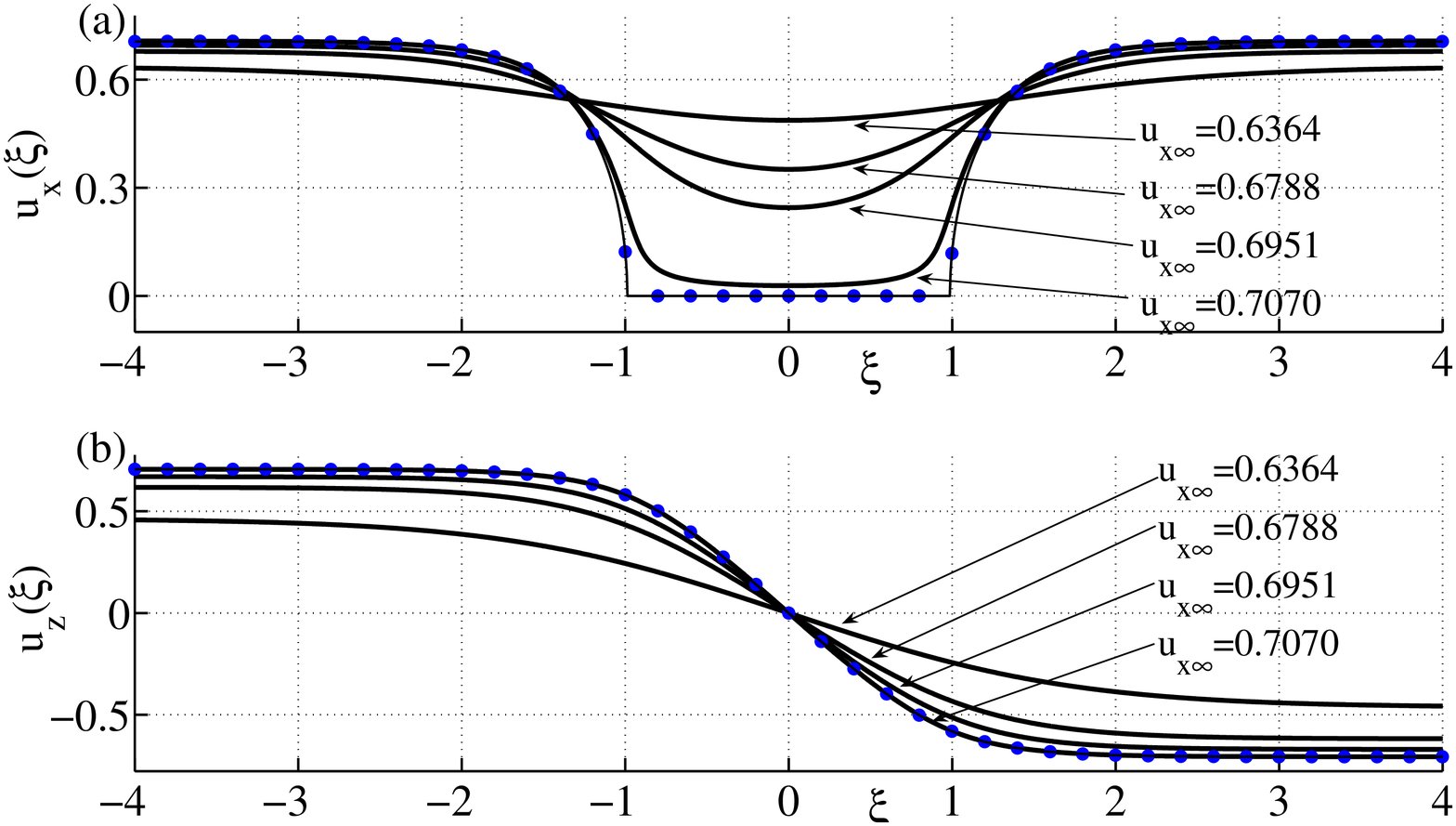}
\caption{(Color online) Profiles of $u_x(\xi)$ (a) and $u_z(\xi)$ (b) of various nonlinear guided waves for $\gamma=-2$ and $u_
{x\infty}=0.6364,0.6788,0.6951,0.7070$ (solid lines). The asymptotical ($u_{x\infty} \rightarrow \sqrt{1/2}$) profiles are also reported (dotted
line).} \label{fig1}
\end{center}
\end{figure}

In order to describe the formation of the hollow core wave (reported in Fig.2) we consider the case $u_{x\infty}^2 = 1/2$ where $\beta=0$ (see the
first of Eqs.(\ref{bet})). In the considered case, the first constraint of Eqs.(\ref{constraints}) is violated so that the analysis of this limiting
solutions has to be based on Eqs.(\ref{syst2}). For $\beta = 0$, the first of Eqs.(\ref{syst2}) reduces to $\epsilon_x^{(NL)} u_x=0$ so that
$\epsilon_x^{(NL)}=0$ or $u_x=0$ are the only two possibilities allowed at each $\xi$. Correspondingly, the second of Eqs.(\ref{syst2}) (for
$\beta=0$), if $\epsilon_x^{(NL)}=0$, yields
\begin{equation} \label{2bet=0}
\frac{d^2 u_z}{ d\xi^2} + \frac{2\gamma}{1+\gamma}(1 -2 u_z^2)u_z =0,
\end{equation}
where use has been made of the relation (see the first of Eqs.(\ref{epNL}))
\begin{equation} \label{uxsho}
u_x (\xi) = \sqrt{\frac{1-(1-\gamma) u_z^2}{1+\gamma}}
\end{equation}
if $u_z^2 > 1/(1-\gamma)$ (assuring the reality of $u_x$ since here $\gamma < -1$). On the other hand, if $u_x=0$, the second of Eqs.(\ref{syst2})
yields
\begin{equation} \label{1bet=0}
\frac{d^2 u_z}{ d \xi^2} + [ 1 -(1+\gamma)u_z^2 ]u_z=0.
\end{equation}
Since asymptotically $u_{x\infty} \neq 0$, the wave has to be asymptotically characterized by the condition $\epsilon_x^{(NL)}=0$ so that, for
positive and large $\xi$, $u_z$ satisfies Eq.(\ref{2bet=0}) whose solution (fulfilling the asymptotical requirements $u_z(+\infty) = - \sqrt{1/2}$
and $[d u_z /d\xi]_{\xi=+\infty}=0$) is
\begin{equation} \label{uzsho}
u_z(\xi) = - \sqrt{\frac{1}{2}} \tanh \left[ \sqrt{\frac{\gamma}{1+\gamma}} (\xi-\xi_0) \right]
\end{equation}
where $\xi_0$ is a real constant.
\begin{figure}[htbp]
\begin{center}
\includegraphics[width=0.5\linewidth]{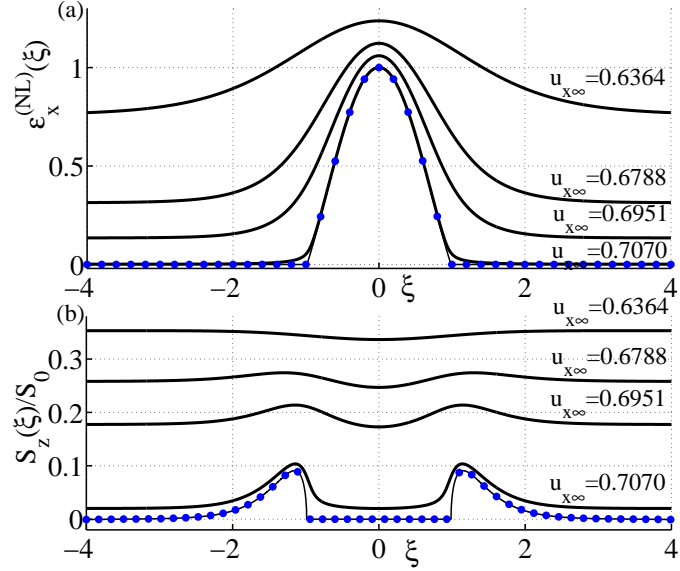}
\caption{(a) (Color online) Profiles of the normalized effective nonlinear dielectric permittivity $\epsilon_x^{(NL)}(\xi)$ supporting the nonlinear
guided waves reported in Fig.2. (b) (Color online) Profiles of the normalized Poynting vector $S_z(\xi)/S_0$ associated with the nonlinear guided
waves reported in Fig.1.} \label{fig3}
\end{center}
\end{figure}
Since $u_z$ of Eq.(\ref{uzsho}) monotonically decreases as $\xi$ decreases, it is evident from Eq.(\ref{uxsho}) that also $u_x$ decreases as $\xi$
decreases and consequently a point $\xi=\Delta$ exists at which $u_x (\Delta)=0$. For $\xi < \Delta$, the profile of $u_z$ of Eq.(\ref{uzsho}) would
yield a complex $u_x$ profile through Eq.(\ref{uxsho}) and this is incompatible with the present discussion. Therefore for $\xi <\Delta$, the
condition $\epsilon_x^{(NL)}=0$ can not consistently hold so that, in this region, the condition $u_x=0$ is the only one left and the field $u_z$ is
locally the solution of Eq.(\ref{1bet=0}) continuously joined (together with its derivative) with Eq.(\ref{uzsho}) at $\xi=\Delta$. We have
numerically evaluated the profile of $u_z$ and $u_x$ in the above discussed case $\beta=0$ and they are reported in Fig.2 (dotted lines) from which
it is evident that the limiting case accurately describes the nonlinear guided waves for $u_{x\infty}>0.7070$ (except for a small region near
$\xi=\pm \Delta$).

From a physical point of view, the fundamental relation $\epsilon_x^{(NL)}u_x=0$ characterizing the above discussed hollow-core waves amounts to the
vanishing of the transverse component of the displacement field, i.e. $D_x=0$, a physical condition differently attained in the core region (where
$E_x=0$) and at the lateral sides (where $\epsilon_x^{(NL)}=0$) of the hollow-core wave, as discussed above. In Fig.3(a), we plot the profiles of
$\epsilon_x^{(NL)}$ associated with the nonlinear guided waves reported in Fig.2 from which we note that, the closer $u_{x\infty}^2$ to $1/2$, the closer
$\epsilon_x^{(NL)}$ to zero in the external regions (i.e. $-\infty<\xi<-\Delta$ and $\Delta<\xi<+\infty$), thus proving the above discussed mechanism
supporting the hollow-core wave. It is worth noting that the relation $\epsilon_x^{(NL)}=0$ can be satisfied only if $|E_{x}|\sim \sqrt{\epsilon/|\chi|}$
and $|E_{z}|\sim\sqrt{\epsilon/|\chi|}$, so that the occurrence of discontinuities in the field spatial derivatives is a signature of the extreme
nonlinear regime where the nonlinear term appearing in the first of Eqs.(\ref{H-D}) is comparable with $\epsilon_0 \epsilon \bf E$, the linear part of
the displacement field. The time averaged Poynting vector ${\bf S}=(1/2)Re[{\bf E} \times {\bf H}^*]$, for the considered fields, is given by ${\bf S}=
S_0 \left(\beta- du_z/d\xi \right) u_x \hat{\bf e}_z$, where $S_0=\sqrt{(\epsilon_0 \epsilon^3)/(4 \mu_0 |\chi|^2)}$ and we report, in Fig.3(b), the
profiles of $S_z(\xi)/S_0$ corresponding to the nonlinear guided waves plotted in Fig.2. Note that, in the limiting case ($\beta=0$, plotted in dotted
line in Fig.3(b)), the energy flow exactly vanishes in the core region $-\Delta<\xi<\Delta$ and it shows very steep slopes at $\xi=\pm \Delta$ (following
the behavior of the $x$-component of the electric field). In addition the power flow is transversally confined since $S_z \rightarrow 0$ for $\xi
\rightarrow \pm \infty$ as a consequence of the fact that, asymptotically, $du_z/d\xi \rightarrow 0$.

\section{Conclusion}

In conclusion, we have investigated the formation of optical hollow-core waves in Epsilon-Near-Zero metamaterials. The existence of hollow-core waves
characterized by a power flow exactly vanishing at a central region is the signature of the extreme nonlinear regime where the linear and nonlinear
contributions to the overall medium dielectric response are comparable.

\end{document}